\documentclass[fleqn,usenatbib]{mnras}
\usepackage{newtxtext,newtxmath}
\usepackage{graphicx}	% Including figure files
\usepackage{amsmath}	% Advanced maths commands
\usepackage{amssymb}	% Extra maths symbols
\usepackage{tabularx}
\usepackage[flushleft]{threeparttable}
\usepackage[a4paper]{geometry}%
\usepackage{hyperref}

\newcommand{\Msun}{~M_\odot}

\title[Thermal conduction in AGN feedback]{Sound wave generation by a spherically symmetric outburst and AGN feedback in galaxy clusters II: impact of thermal conduction.}

\author[Xiaping Tang, Eugene Churazov]{
Xiaping Tang,$^{1,2}$\thanks{E-mail: tangxiaping@gmail.com, tang.xiaping@mail.huji.ac.il}
Eugene Churazov,$^{1,3}$
\\
$^{1}$Max Planck Institute for Astrophysics,
Karl-Schwarzschild-Str. 1,
D-85741 Garching, Germany\\
$^{2}$ The Racah Institute of physics, The Hebrew University of Jerusalem, Jerusalem 91904, Israel\\
$^{3}$ Space Research Institute, Profsoyuznaya str. 84/32, Moscow
  117997, Russia\\
  }

\date{Accepted XXX. Received YYY; in original form ZZZ}

\pubyear{2018}

\begin{document}
\label{firstpage}
\pagerange{\pageref{firstpage}--\pageref{lastpage}}
\maketitle

\begin{abstract}
We analyze the impact of thermal conduction on the appearance of a shock-heated gas shell which is produced when a spherically symmetric outburst of a supermassive black hole inflates bubbles of relativistic plasma at the center of a galaxy cluster. The presence of the hot and low-density shell can be used as an ancillary indicator for a high rate of energy release during the outburst, which is required to drive strong shocks into the gas. Here we show that  conduction can effectively erase such shell, unless the diffusion of electrons is heavily suppressed. We conclude that a more robust proxy to the energy release rate is  the ratio between the shock radius and bubble radius. We also revisited the issue of sound waves dissipation induced by thermal conduction in a scenario, where characteristic wavelength of the sound wave is set by the total energy of the outburst. For a fiducial short outburst model, the  dissipation length does not exceed the cooling radius in a typical cluster, provided that the conduction is suppressed by a factor not larger than $\sim$100.  For quasi-continuous energy injection neither the shock-heated shell nor the outgoing sound wave are important and the role of conduction is subdominant. 
\end{abstract}

\begin{keywords}
conduction -- shock waves -- galaxies: active -- galaxies: clusters: individual:
M87 -- galaxies: clusters: individual: Perseus cluster -- galaxies: clusters: intracluster medium.
\end{keywords}

\section{Introduction}
Supermassive black holes (SMBHs) could be an important source of energy for the gas in forming galaxies \citep[e.g.][]{1998A&A...331L...1S}, which prevents excessive star formation, particularly at the high-mass end of the galaxies distribution. However, the first reliable evidence that SMBHs do affect the gas came from radio and X-ray observations of galaxy clusters \citep[e.g.][]{1993MNRAS.264L..25B,1995MNRAS.274L..67B} that revealed bubbles of radio-bright plasma inflated in the intracluster medium (ICM).  The clarity of AGN Feedback features in clusters is driven by two factors i) the scale of the energy  injection by SMBHs in these objects is larger than in individual galaxies and ii) the AGN Feedback in clusters continues to be important in the Local Universe. A lower limit on the energy injection rate can be estimated by comparing the bubble's inflation time scale with the buoyancy time scale \citep[][see also Pedlar et al., 1990; Binney \& Tabor, 1995]{2000A&A...356..788C,2000ApJ...534L.135M}.  These estimates have shown that in few best studied clusters the mechanical power of an AGN matches gas cooling losses and exceeds by far the radiative power of these AGNs \citep[e.g.][]{2001ApJ...554..261C,2002A&A...382..804B}. Studies of samples of clusters confirmed this conclusion \citep[e.g.][]{2004ApJ...607..800B}. Furthermore, the energy conservation arguments suggest that much of the mechanical power will be transfered to the ICM during buoyant rise of the bubble \citep[][]{2001ApJ...554..261C,2002MNRAS.332..729C,2001ASPC..250..443B}  implying a high efficiency of the feedback and a possibility that the AGN power adjusts to match gas cooling losses. 

Of course, a comprehensive  AGN Feedback model should include a detailed description of how the gas gets to the black hole, the accretion physics and the jet formation, as well as physical processes that lead to the dissipation of the released energy in the ICM, etc \citep[see, e.g.,][for recent reviews of different scenarios]{2012ARA&A..50..455F,2012NJPh...14e5023M,2014PhyU...57..317V,2016NewAR..75....1S}.
Here we address one element of the above problem, namely, the role of the thermal conduction in the scenario involving inflation of bubbles by a SMBH. This problem is particularly interesting in connection with the possible role of sound waves in the ICM heating. The important role of sound waves is advocated in, e.g., \citet[][]{Fabian05,2017MNRAS.464L...1F,2004ApJ...611..158R,FS05,2009MNRAS.392.1413S,2009MNRAS.395..228S}. On the other hand, in the buoyancy-driven feedback scenario \citep[e.g.][]{2000A&A...356..788C,2016MNRAS.458.2902Z,2017ApJ...844..122F} the role of the sound waves in the energy budget is subdominant to the enthalpy of the inflated bubbles, which release energy as they rise through the stratified ICM atmosphere. For instance, for the case of a spherically symmetric outburst in a homogeneous medium, the amount of energy associated with the outgoing sound wave is only a small portion $\zeta_{wave}$ of the total injected energy (Tang\& Churazov. 2017, hereafter TC17). $\zeta_{wave}$ depends on the duration time of an outburst and reaches maximum for an instantaneous outburst, which drives a strong shock into the ICM. In this case, a hot low-density shell is formed by the shock-heated gas. The thermal conduction could affect both the appearance of this hot shell and the rate at which outgoing sound wave is losing its energy.

Assuming the injected materials has the same adiabatic index as the ICM for simplification, i.e. $\gamma_{ej}=\gamma_{ICM}=5/3$, \cite{TC17} found that for an instantaneous outburst $\zeta_{wave}$ does not exceed $\sim$12.5\%. The injected materials from the central AGN is plausibly a purely relativistic plasma or a mixture of relativistic plasma and thermal gas. The corresponding adiabatic index of ejecta $\gamma_{ej}$ varies between $4/3$ and $5/3$. If $\gamma_{ej}=4/3$, the fraction of energy carried out by sound wave $\zeta_{wave}$ becomes even smaller, as larger thermal energy is associated with the bubble of a given size. This, however, is important only for the outbursts, which are not too short, since otherwise the size of the bubble is negligible. 

The heat flux driven by the global radial temperature gradient in cool-core clusters has been considered as one of the possible ways to compensate for the gas cooling losses \citep[e.g.][]{1983ApJ...267..547T,1986ApJ...306L...1B,2002ApJ...581..223R,2003ApJ...582..162Z,2004MNRAS.347.1130V,2008ApJ...688..859G}. Here we ignore the effects of the global heat flow by considering the outburst in the initially isothermal gas.

The structure of this paper is as follows. In \S\ref{sec:basic} we describe our spherically symmetric outburst model and basic assumptions. In \S\ref{sec:shell} and \ref{sec:sound}, we study the impact of thermal conduction on the shock-heated shell and the dissipation of sound waves induced by the outburst, respectively. In \S\ref{sec:discussion} we discuss the implications in the context of AGN Feedback in galaxy clusters.

\section{Basic model}
\label{sec:basic}
In this section we outline basic assumptions made in order to simplify the problem. 
\begin{figure}
 \begin{center}
 \includegraphics[width=\columnwidth]{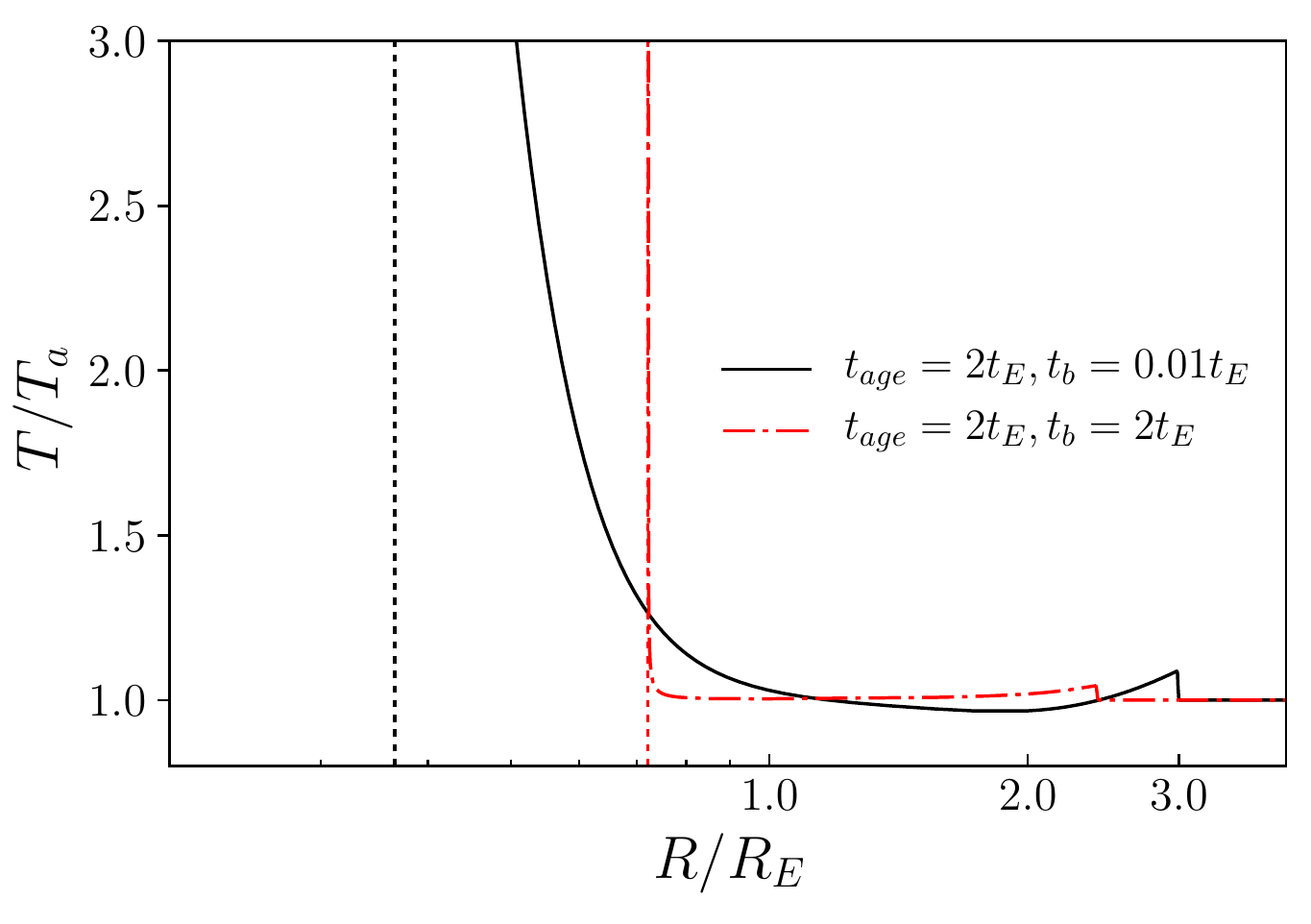} 
  \caption{Comparison of temperature distributions at $t_{age}=2t_E$ for a fast/short outburst ($t_b/t_E=0.01$) and a slow/long outburst ($t_b=2t_E$) in uniform medium without thermal conduction (from \citeauthor{TC17}). Two vertical dotted lines show the ejecta boundaries for $t_b=0.01t_E$ (black) and $t_b=2t_E$ (red) respectively. For the slow outburst (red dotted-dashed line), the amplitude of the outgoing wave is smaller and the ICM temperature in the inner region (but outside the ejecta boundary) is close to the initial value $T_a$. In this case the impact of thermal conduction is minimal. The opposite is true for the short outburst (black solid line), which develops large temperature variations across the radius. }
    \label{fig:various_tb}
 \end{center}
 \end{figure} 
 
\subsection{Spherically symmetric outburst}
\label{sec:sph}
We limit our discussion to a spherically symmetric outburst in homogeneous medium. As we have shown in \citeauthor{TC17}, in the energy-dominated (rather than momentum-dominated) feedback scenario such setup is suitable for the environment in the cores of clusters with relatively shallow density/pressure profiles. The medium has an initial pressure $P_a$, density $\rho_a$, and temperature $\displaystyle T_a=\frac{P_a\mu m_p}{\rho_a k_B}$, where $m_p$ is the proton mass, $\mu$ is the mean atomic weight and $k_B$ is the Boltzmann constant. During the outburst an AGN releases energy and matter in a small region near the center. We assume the injected materials and the ICM have the same adiabatic index $\gamma=5/3$ in the rest of our discussion for simplification. We employed the Lagrangian scheme in the numerical simulations, see Appendix {\ref{app:numerical}} for details. The energy is released in the central cell, which then expands into the surrounding medium (ICM). Below we often refer to this central cell as a ``bubble'' or ``ejecta''. The first name emphasizes the similarity with radio-bright bubbles observed in the cores of relaxed clusters, while the second name referrers to the analogy with the physics of explosions in, e.g., supernova.  As the boundary of the central cell moves, it serves as a piston that drives a perturbation into the ICM. In  \citeauthor{TC17}, see also \cite{2017ApJ...844..122F}, we characterize the outburst with two parameters: total energy $E$ released during the outburst and the duration of the outburst $t_b$. The value of $E$ defines the characteristic length scale $R_E$
\begin{equation}
R_{E}\sim \left(\frac{3E}{4\pi P_a}\right)^{1/3}\sim15\,{\rm kpc}\,E_{59}^{1/3}n_{e,-2}^{-1/3}T_{a,8}^{-1/3},
\label{eq:R_E}
\end{equation}
where $\displaystyle E_{59}=E/10^{59}~{\rm erg}$, $T_{a,8}=T_a/10^8~{\rm K}$, $n_{e,-2}=n_e/10^{-2}~{\rm cm^{-3}}$. The corresponding time scale $t_{E}$ is 
\begin{equation}
t_{E}\sim R_{E}/c_s=9.4\, {\rm Myr}\, E_{59}^{1/3}n_{e,-2}^{-1/3}T_{a,8}^{-5/6},
\end{equation}
where $\displaystyle c_s=\sqrt{\gamma P_a/\rho_a}$ is the adiabatic sound speed. 

\begin{figure}
 \begin{center}
 \includegraphics[width=\columnwidth]{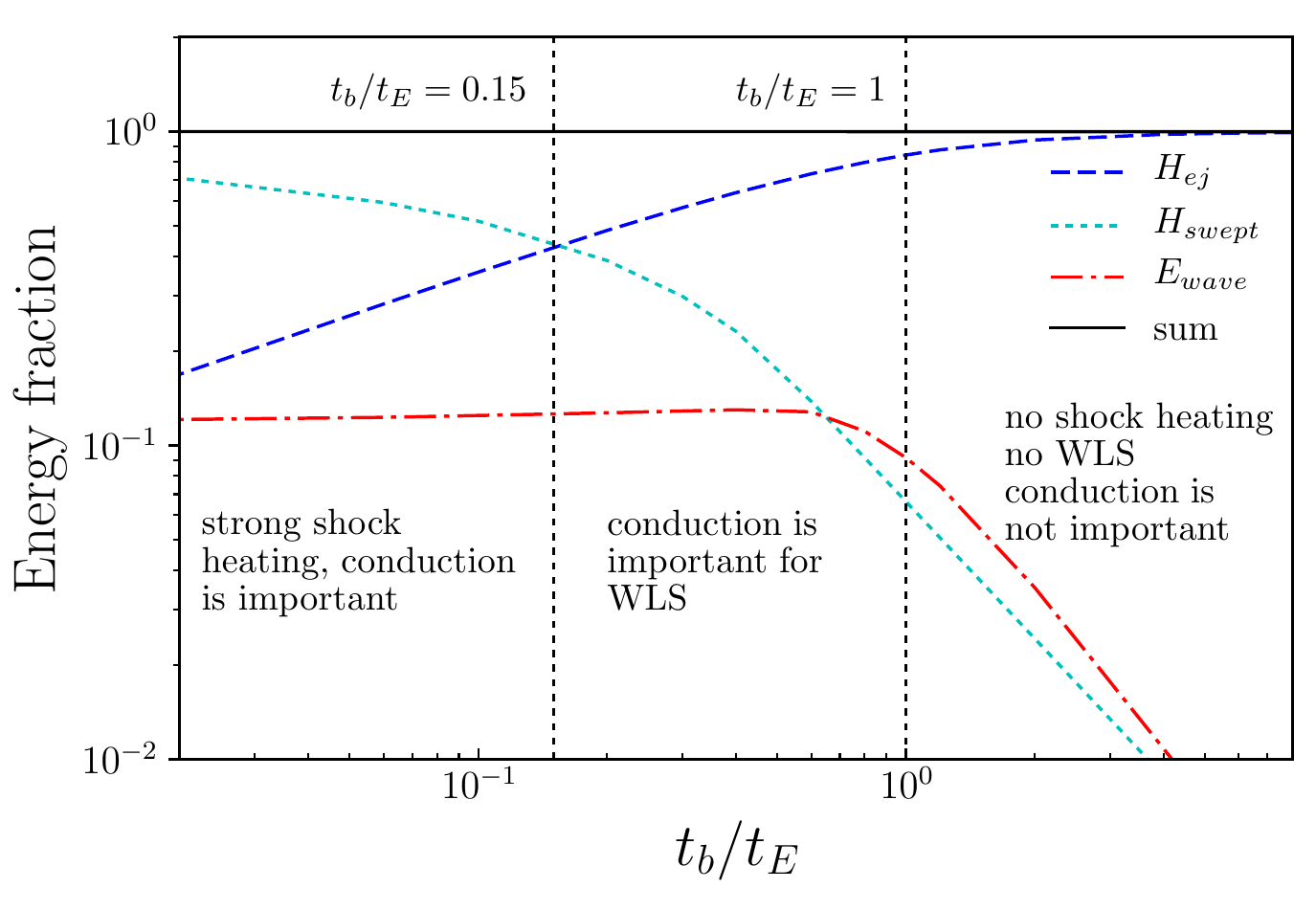} 
  \caption{Energy partition of an outburst as a function of $t_b/t_E$ ratio. $H_{ej}$ is the enthalpy of ejecta, $E_{wave}$ is the energy carried away by the wave-like structure (outgoing sound wave) and $H_{swept}$ represents the energy available for radiative cooling in the shocked materials, see \citeauthor{TC17} for details. The partition of energy is evaluated at $t_{age}/t_E=10$, which could be considered as a good approximation for the asymptotic limit at $t_{age}/t_E\rightarrow \infty$. Two vertical dotted lines schematically divide the outbursts according to the $t_b/t_E$ ratio into three regimes. When  $t_b/t_E\lesssim 0.15$, the shock-heated gas shell captures most of the outburst energy and an outgoing sound wave carries $\sim 10$\% of energy. In the intermediate regime $1 \gtrsim t_b/t_E\gtrsim 0.15$, the enthalpy of the bubble dominates, while the outgoing sound wave still carries $\sim 10$\% of energy. For $t_b/t_E\gtrsim 1$ vast majority of energy goes into the bubble enthalpy, while the contributions of the shock-heated shell and the sound wave are small. }
    \label{fig:energy_partition}
 \end{center}
 \end{figure}

For the purpose of this paper, we divide the outbursts into two groups according to the ratio $t_b/t_E$. The limit of $t_b/t_E\ll1$ corresponds to an instantaneous outburst, i.e. the classical Sedov-Taylor (ST) solution \citep{Taylor46,Sedov59}, in which a strong shock is driven into the ICM. As a result, at time $t\gg t_E\gg t_b$  a hot and low density shell is present in the central region, surrounding the ejecta, as well as an outgoing sound wave, carrying $\sim 12.5$\% of total outburst energy $E$ [see, e.g., \citeauthor{TC17}; \cite{2017ApJ...844..122F}]. In the opposite limit of $t_b/t_E \gg 1$, the expansion of the ejecta boundary is subsonic and the boundary merely doing $PdV$ work on the gas, displacing it to larger radii, while the most significant part of energy remains in the form of thermal energy of the ejecta. In this case, no shock-heated shell is formed and the amount of energy carried by the sound wave is small. A comparison of the temperature profiles for the above-mentioned limits, when conduction is ignored, is shown in Fig.~\ref{fig:various_tb}. The energy partition of an outburst for different $t_b/t_E$ ratios is presented in Fig. \ref{fig:energy_partition}. Thermal conduction is expected to be only important in fast/short outbursts, in which a shock-heated shell surrounding the ejecta exists and the sound wave also carries a significant fraction of outburst energy. As a result, in the following discussion we focus on the instantaneous outburst to study the effect of thermal conduction.

\subsection{Thermal conduction}
For simplicity, we assume that the electron and ion temperatures are equal ($T=T_e=T_i$) and the heat flux $Q$ is given by 
\begin{eqnarray}
Q=-\kappa \nabla T,
\label{eq:j}
\end{eqnarray}
where $\kappa$ is the thermal conductivity. We adopt the parameterization of $\kappa$ in \cite{CM77,1982ApJ...252..529B} 
\begin{equation}
\kappa=\xi\kappa_{sp}\left(1+\frac{\xi}{\xi_{sat}} 4.2 \lambda_e |\nabla \ln T|\right)^{-1},
\label{eq:kappa}
\end{equation}
to include the effect of saturation. $\kappa_{sp}\approx 1.3 n_e \lambda_e k_B \left ( \frac{k_BT}{m_e}\right )^{1/2} \approx 5\times 10^{13}\, T_8^{5/2}\rm erg\, s^{-1}cm^{-1} K^{-1}$ is the Spizter conductivity \citep{Spitzer56}; $\displaystyle \lambda_e=2.3~{\rm kpc}~ T_8^2 n_{e,-2}^{-1}$ is the electron mean free path for Coulomb collisions and $\xi$ is the suppression of the conductivity relative to the Spitzer value, equivalent to the assumption that the effective mean free path is $\xi \lambda_e$. The parameter $\xi_{sat}$ in eq. \ref{eq:kappa} controls the level of the saturated heat flux. $\xi_{sat} \rightarrow \infty$ implies unsaturated heat flux, while $\xi_{sat}=1$ corresponds to the case considered by \cite{CM77}. We note here that for a weakly collisional high-$\beta$ plasma (where $\beta$ is the ratio of thermal pressure to magnetic pressure) a comprehensive description of heat fluxes is an open issue and it may require much more complicated recipes than those provided by eqs. (\ref{eq:j}) and  (\ref{eq:kappa}) \citep[see, e.g.][]{1998PhRvL..80.3077C,2001ApJ...549..402M,
2001ApJ...562L.129N,2016MNRAS.460..467K,RC16,2017arXiv171111462K,RC18}. 
However, setting $\xi=1$ and $\xi_{sat}\sim 1$ plausibly corresponds to the case, when the impact of thermal conduction is maximal. We also emphasize that the heat fluxes through the boundary of the bubble/ejecta are assumed to be zero and the thermal conductivity operates only in the ICM outside the ejecta boundary.

Previous studies of thermal conduction in spherically symmetric outbursts focus on the early time evolution, when the shock is still strong. For instance, a self-similar solution for the conduction-driven thermal wave is derived  \citep[see, e.g.][]{Barenblatt96,RM91}, see Appendix \ref{appendix:TCD} for details. Numerical efforts are also made in, e.g., the context of supernova remnant evolution \citep{Chevalier75}.  Here we are interested in the late time evolution in the ICM environment. We emphasize that we ignore any temperature gradients in the initial temperature distribution. While this is not consistent with the observed temperature gradients in the cool-core clusters, these gradients are typically smaller than the ones generated by the outburst.  

According to the discussion in \S\ref{sec:sph} (see also Fig.~\ref{fig:various_tb}), in the case of $t_b\gg t_E$, i.e., ``slow'' outburst, neither the hot shock-heated shell nor the sound wave carrying significant amount of energy are formed. Thus we do not expect the thermal conductivity to make any substantial impact in this case.  We therefore focus the discussion below on the case  $t_b/t_E\ll1$, i.e., an instantaneous outburst, where the impact of thermal conduction is expected to be much more significant. In particular, we want to address the following two major questions:  
\begin{enumerate}
\item How does thermal conduction affect the properties of the strongly shock-heated gas shell? 
\item How quickly the energy of the outgoing sound wave is dissipated? 
\end{enumerate}

\section{Shock-heated gas shell}
\label{sec:shell}

\begin{figure*}
 \begin{center}
 \includegraphics[width=0.9\textwidth]{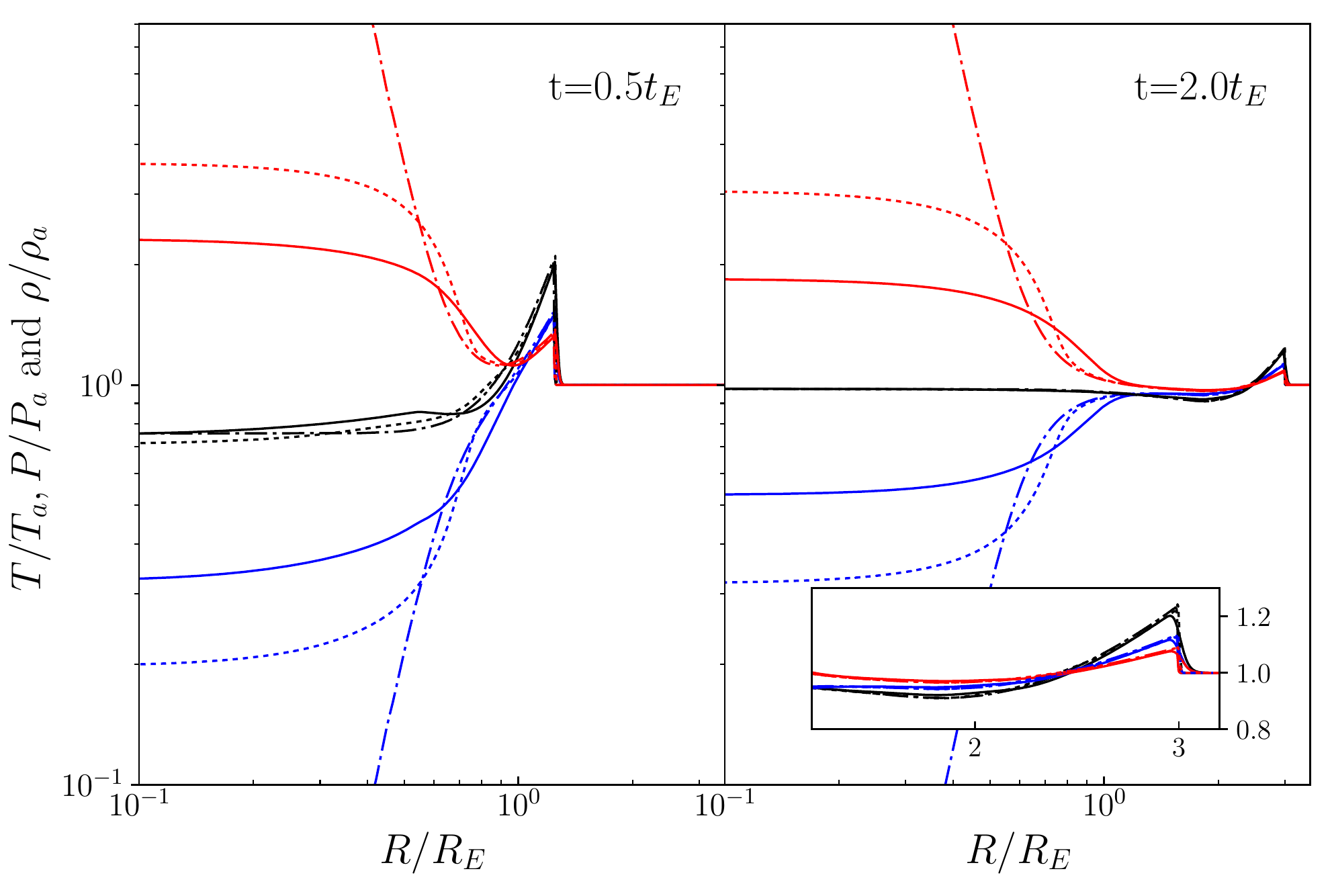} 
  \caption{Impact of thermal conduction on the normalized temperature $T/T_a$ (red), pressure $P/P_a$ (black) and density $\rho/\rho_a$ (blue) profiles, plotted as a function of dimensionless radius $R/R_{E}$. The left and right panels correspond to $t=0.5t_E$ and $2\;t_E$, respectively.   The solid line, dotted and dotted dashed lines are for $\xi=1$, $0.1$ and $0.01$ respectively. The inlay in the right panel is a zoom-in view of the shock and the wave like structure downstream of the shock front in linear scale. It is clear that conduction has a very strong impact on the inner shock-heated region, while the shock is only slightly affected at $t\lesssim 2 t_E$.}
    \label{fig:inner_structure}
 \end{center}
 \end{figure*} 
 
An outburst with $t_b\ll t_E$ always generates a shell of shock-heated gas, which captures the dominant fraction of the outburst energy, see the cyan line in Fig. \ref{fig:energy_partition}. At time $t\gg t_E \gg t_b$ the pressure in this shell approaches the initial pressure and the gas inside the shell is hot and low density, making the thermal conduction especially effective. 
For an instantaneous outburst without thermal conduction in a homogeneous gas with known thermodynamic properties, the only relevant burst parameter is the total energy release $E$, which sets the characteristic length scale $R_E$ and time scale $t_E$. It is instructive to compare $t_E$ with the characteristic conduction time $t_c$ at $t=t_E$, ignoring effects of saturation. At $t\sim t_E$, the shock is approximately at radius $\sim R_E$, the total thermal energy of the swept gas is comparable to $E$ and the temperature of the gas behind the shock is of the order of $T_a$. The characteristic conduction time scale can be evaluated as the ratio of the radius squared to the conductivity/diffusivity of electrons, i.e.,  
\begin{eqnarray}  
t_c(R_E)=\frac{n_e R_E^2 k_B}{\kappa}\approx \frac{R_E^2}{1.3 \xi \lambda_e  \left ( \frac{kT}{m_e}\right )^{1/2}},
\end{eqnarray}  
where $T\sim T_a$ and $\lambda_e\sim \lambda_e(T_a)$.  The ratio of the conduction time $t_c(R_E)$  to $t_E$ is 
\begin{eqnarray}  
\theta_{c,E}=\frac{t_c(R_E)}{t_E}\approx \frac{R_E c_s}{1.3 \xi \lambda_e  \left ( \frac{kT}{m_e}\right )^{1/2}}\approx \frac{R_E}{43 \xi \lambda_e }.
\end{eqnarray} 
Based on above calculation, if $R_E$ is smaller than $\sim 43 \xi \lambda_e$, then conduction can erase the temperature gradient associated with the shock-heated gas shell at $t\sim t_E$. We can extend this analysis to earlier time evolution of an outburst. In the supersonic expansion phase, the characteristic gas velocity ${\rm v}$ can be estimated through the relation $\displaystyle E\sim 2\pi\rho {\rm v}^2 R^3/3$, where $R$ is the current shock radius and $k_BT\sim m_p{\rm v}^2$. Given that $\lambda_e\propto T^2\propto {\rm v}^4 \propto R^{-6}$ (here we only keep the temperature dependence in the expression of $\lambda_e$), the ratio $t_c(R)/t$, where $t=R/{\rm v}$, is a strong function of $R$, namely 
 \begin{eqnarray}  
\frac{t_c(R)}{t}\approx \theta_{c,E} \left ( \frac{R}{R_E} \right ) ^7.
\end{eqnarray} 
Therefore, during the initial phase of the outburst the conduction time can be shorter than the expansion time. If $\theta_{c,E}\ll 1$ then the traces of the shock-heated gas will be promptly erased at all phases of the outburst evolution. If $\theta_{c,E}\gg 1$ then traces of the shock-heated gas shell will persist long after the characteristic gas velocities become subsonic.

\begin{figure}
 \begin{center}
 \includegraphics[width=\columnwidth]{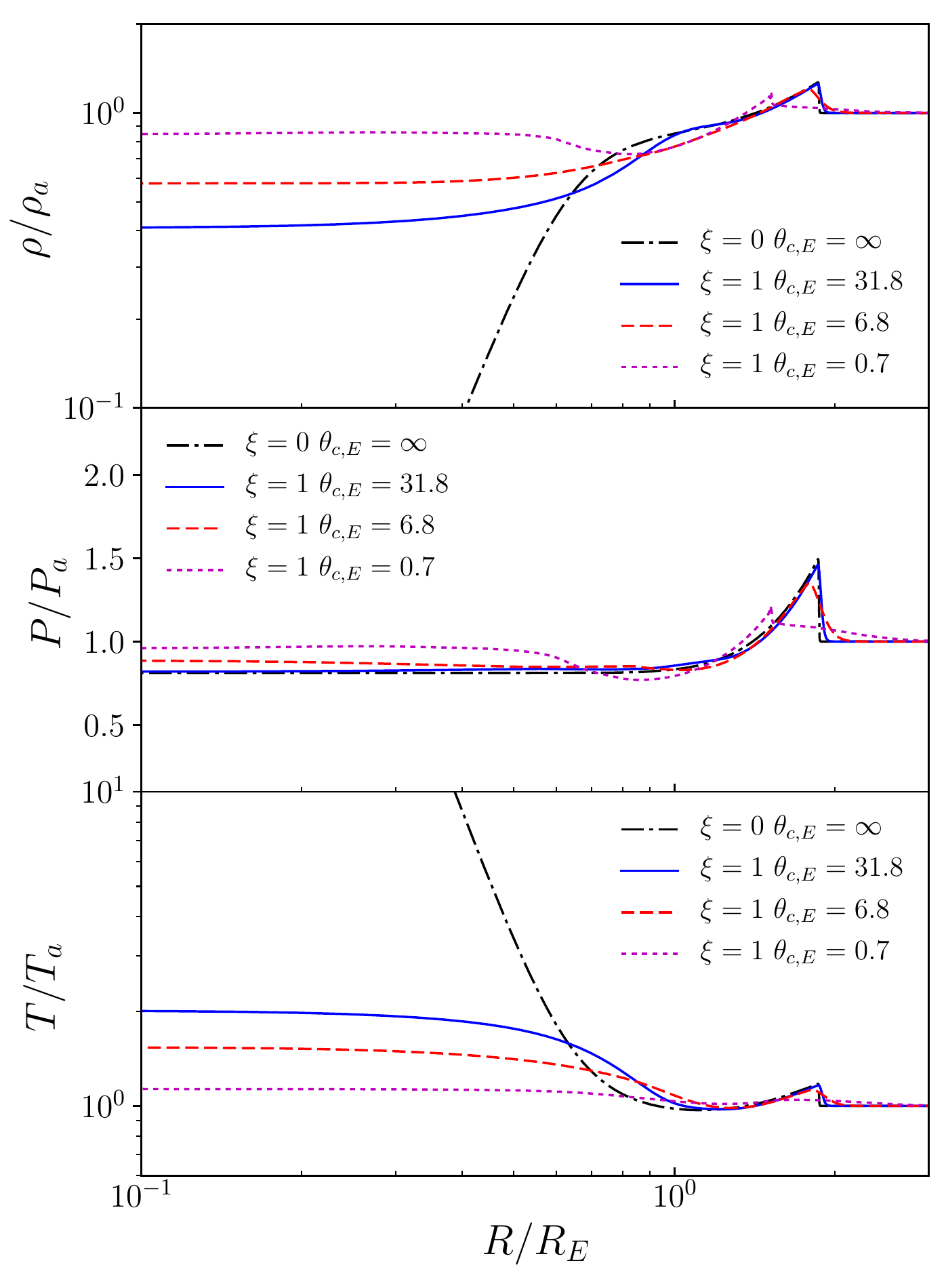} 
  \caption{Impact of the parameter $\theta_{c,E}$ on the density $\rho/\rho_a$, pressure $P/P_a$ and  temperature $T/T_a$ distribution at $t=t_E$. In these runs we vary $E_{59}$, $T_{8}$ and $n_{e,-2}$, resulting in variations of $\theta_{c,E}$. The smaller is the value of $\theta_{c,E}$, the larger are the deviations from the run, which ignores the conduction. Once $\theta_{c,E}\lesssim 1$, the changes in the front become noticeable, the outgoing wave becomes almost isothermal, has smaller amplitude and sound speed.}
    \label{fig:T_n}
 \end{center}
 \end{figure}

The above consideration provides only qualitative estimates and completely ignores the saturation of heat fluxes. A more accurate treatment of the conductivity [in the form described by eq.~(\ref{eq:kappa})] can be done in numerical simulations. A fiducial run with different $\xi$ is shown in Fig.~\ref{fig:inner_structure}. One can see that the change of $\xi$ from 0.01 to 1 produces a significant drop in the characteristic amplitude of temperature variations in the shock-heated shell. The value of  $\xi_{sat}$ is fixed to 1 in all these runs. The presence of conduction breaks the self-similarity of the instantaneous outburst, by adding one extra parameters $\kappa$, or equivalently  $\theta_{c,E}$. A sample of runs with different  $\theta_{c,E}$ are shown in Fig.~\ref{fig:T_n}.  The smaller is the value of $\theta_{c,E}$, the larger are the deviations from the run without conduction. It is interesting to note that as $\theta_{c,E}$ decreases below $\sim 1$, the adiabatic sound wave gradually becomes isothermal. Because isothermal sound speed is smaller, the wave-like structure in the run with $\theta_{c,E}=0.7$ lags behind the similar structure in the runs with larger $\theta_{c,E}$.

\begin{figure}
 \begin{center}
 \includegraphics[width=\columnwidth]{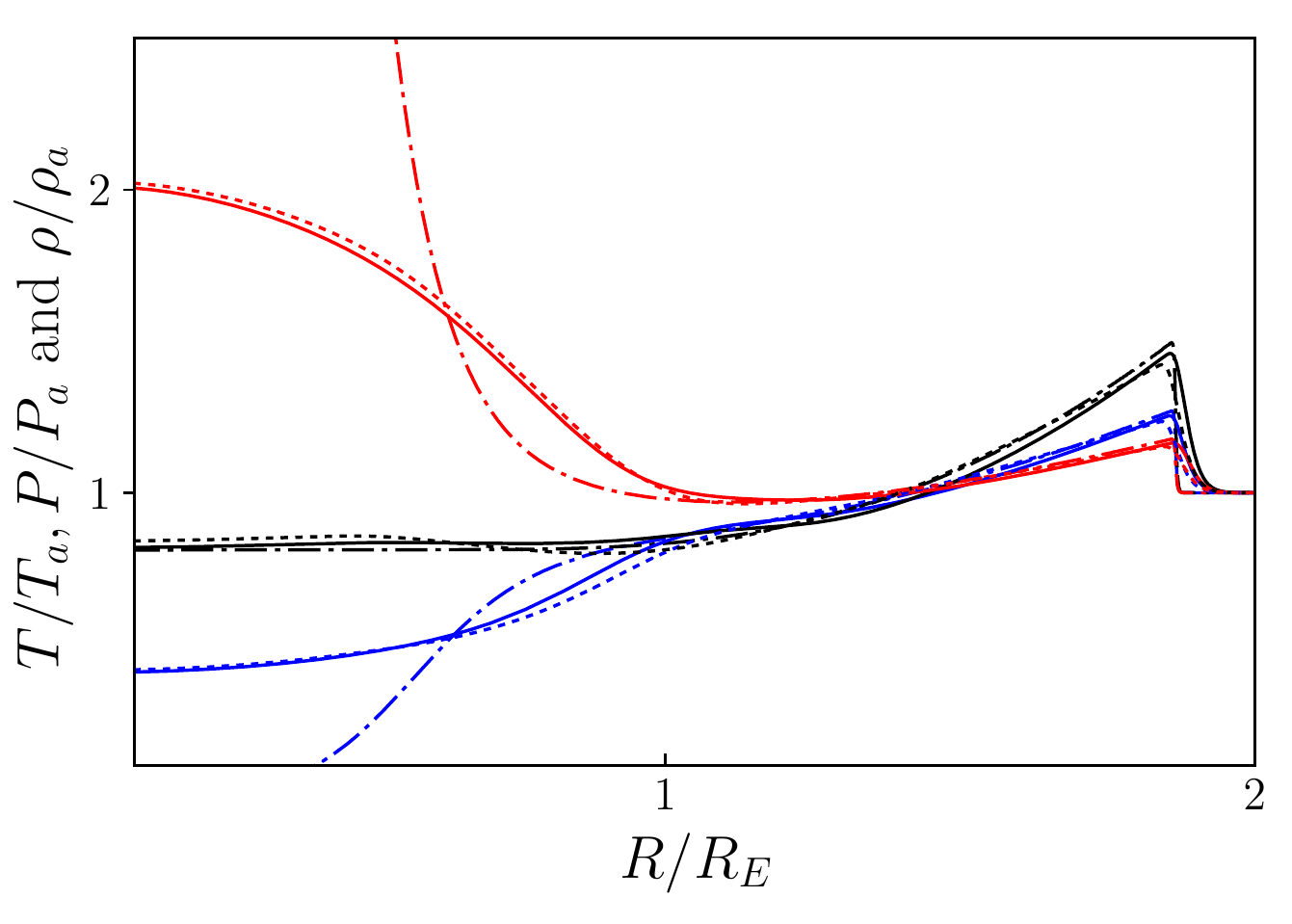} 
  \caption{Effect of saturated heat flux on the $T/T_a$ (red), pressure $P/P_a$ (black) and density $\rho/\rho_a$ (blue)  profiles at $t=t_E$. The dotted-dashed line represents the case without thermal conduction. The solid and dotted lines are for $\xi_{sat}=1$ and $10$, respectively. $\xi_{sat}=1$ corresponds to the case considered in Cowie \& McKee (1977). For $\xi_{sat}=10$ the saturated flux is 10 times higher. Despite the different levels of saturation the resulting profiles are similar, implying that at $t=t_E$, and hence at $t>t_E$, the effects of saturation are small.}
    \label{fig:saturation}
 \end{center}
 \end{figure}

Yet another parameter affecting the solution is the level of saturated flux, controlled by the value of $\xi_{sat}$ in eq.~(\ref{eq:kappa}). One can compare the estimated level of the heat flux with the saturated limit at $t\sim t_E$. An estimate of the heat flux is  $Q\sim \xi \kappa_{sp} T_a/R_E$, while the saturation limit is $Q_{sat}=\xi_{sat}\kappa_{sp}T_a/4.2\lambda_e$ \citep{CM77}. Their ratio is 
\begin{equation}
\frac{Q}{Q_{sat}}\sim \frac{\xi}{\xi_{sat}}\frac{4.2\lambda_e}{R_E}\sim \frac{\xi}{\xi_{sat}} 0.7\,E_{59}^{-1/3}n_{e,-2}^{-2/3}T_{a,8}^{7/3},
\label{eq:saturation}
\end{equation}
which is very sensitive to the ambient temperature $T_{a}$.  An illustration of the effect of saturation is given in Fig.~\ref{fig:saturation}. For our fiducial run with $\xi=1$, $\xi_{sat}=1$, $E_{59}=1$, $n_{e,-2=1}$ and $T_8=0.1$, we have $Q/Q_{sat}=0.003$. Therefore, as expected, the saturation has small effect on the profiles at $t=t_E$. However, for earlier phases of the outburst, when $T\gg T_a$ and $R_s\ll R$ the saturation can limit the heat flux.

In Table \ref{tab:Persesu_M87}, we present the $\theta_{c,E}$ and $Q/Q_{sat}$ for the Perseus cluster and M87 with $\xi=1$ and $\xi_{sat}=1$. According to Fig. \ref{fig:T_n} and \ref{fig:saturation}, strong thermal conductivity close to Spitzer value is able to erase the shock-heated shell in Perseus cluster and M87, while the saturation [in the form suggested by \citep{CM77}] is expected to be unimportant. The results presented here agree qualitatively with that of \cite{Graham08}, although quantitatively their conclusion that a very large amount of energy is present in a shock-heated shell is at tension with our earlier models of a continuous outburst tailored for the parameters of the Perseus cluster \citep[see][TC17]{2016MNRAS.458.2902Z}. The difference is at least partly caused by different assumptions of the shock and the bubble radii. Besides, their conclusion is based on the data analysis rather than on a numerical model.

\begin{table}
\centering
\caption{Basic parameters for Perseus cluster and M87}
\begin{tabular}{cccccc}
\hline\hline
object & $E_{59}$& $n_{e,-2}$& $T_8$& $\theta_{c,E}$ & $Q/Q_{sat}$ \\
\hline
Perseus & 1.08 & 10& 0.4 & 6 & 0.017\\
M87 & 0.055 & 4.5 & 0.2 & 6.5 &0.016\\

\hline\hline
\end{tabular} 
\label{tab:Persesu_M87}
\end{table}

\section{Dissipation of sound waves}
\label{sec:sound}
Based on Fig.~\ref{fig:energy_partition}, in the absence of thermal conductivity an instantaneous outburst drives a spherical shock into the surrounding medium. As the shock propagates to larger radii, it becomes weaker and gradually approaches a sound wave, which carries $\sim12.5$\% of the outburst energy at $t\gg t_E$  assuming $\gamma_{ej}=\gamma_{ICM}=5/3$ (see, e.g. \citeauthor{TC17}). The dissipation of sound waves due to the thermal conduction is suggested as a possible mechanism for heating the ICM in the cool-core clusters \citep[e.g.][and references therein]{Fabian05,2017MNRAS.464L...1F}. Here we revisit this problem as a part of the instantaneous outburst model outlined in \S\ref{sec:basic}. 
\begin{figure}
 \begin{center}
 \includegraphics[width=\columnwidth]{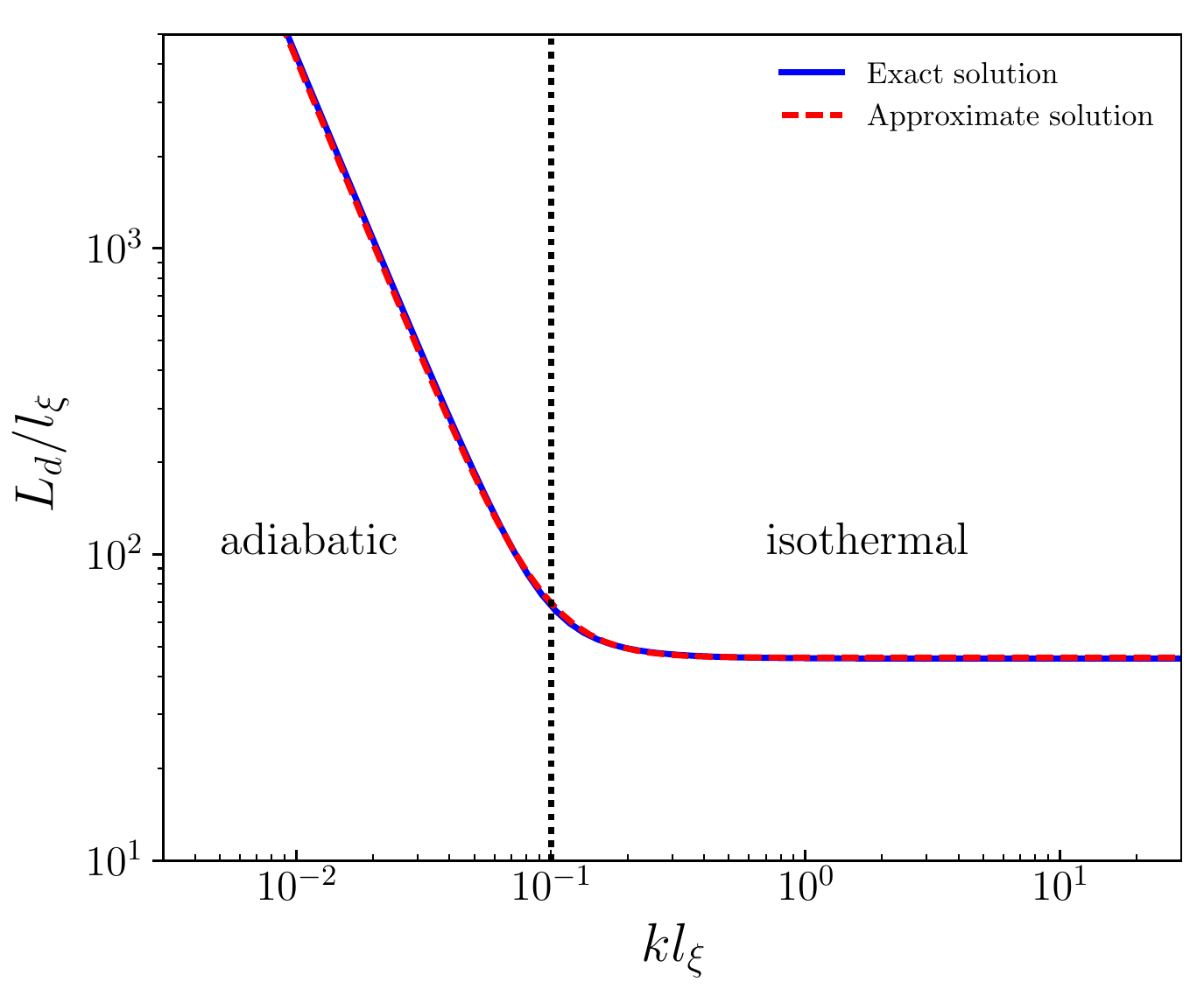} 
  \caption{The decay length $L_d$ of sound waves amplitude in units of $l_\xi$  as a function of $k l_\xi$, where $k$ is the wavenumber and $l_\xi=\xi \lambda_e$ is the effective electron mean free path. The energy decay length is half of $L_d$. The black solid line represents the exact solution, while the red line shows approximation given by eq.~(\ref{eq:lda}). When $k l_\xi\lesssim 0.1$ the wave is adiabatic, while at large $k$ it is isothermal. In the isothermal regime, the decay length is $\approx 46 l_\xi$.} 
\label{fig:DL_dimensionless}
 \end{center}
 \end{figure}

At first, we want to show that saturation is unimportant for estimating the decay rate of a weak shock.  For a weak shock with Mach number $M\sim 1$, electrons can smear the temperature gradient on scales $\displaystyle \Delta x \sim \xi \lambda_e \sqrt{m_p/m_e}\sim 43 \xi \lambda_e$.  As a result, the ratio of the diffusive heat flux to the saturated heat flux becomes
\begin{equation}
\frac{Q}{Q_{sat}}\sim \frac{\xi}{\xi_{sat}}\frac{4.2\lambda_e}{43\xi  \lambda_e}\sim \frac{0.1}{\xi_{sat}} \ll 1.
\label{eq:saturation_s}
\end{equation}
Hence, the effect of saturation is negligible and the decay rate can be estimated from the dispersion relation for sound waves in the gas with non-zero thermal conductivity, which reads
\begin{eqnarray}
\omega^3-\gamma c_{s,i}^2\omega k^2+ik^2 \Sigma \left [ \omega^2- c_{s,i}^2 k^2\right] = 0.
\label{eq:disp}
\end{eqnarray}
$\omega$ and $k$ are the frequency and the wave number, respectively. $\displaystyle  c_{s,i}=\sqrt{\frac{kT}{\mu m_p}}$ is the isothermal sound speed and 
\begin{eqnarray}
\displaystyle \Sigma=\frac{\xi \kappa_{sp}\mu m_p (\gamma-1)}{\rho k_B}\approx \xi \frac{1.31 (\gamma-1)}{\left (n_t/n_e \right)}\lambda_e \sqrt{\frac{k_B T}{m_e}},   
\end{eqnarray}   
where $n_t/n_e\approx 1.9 $ is the ratio between the total and electron particle number densities. 
This equation has three roots, which correspond to the forward and backward propagating modes and a stationary decaying mode. In the limit of small wave numbers $k\rightarrow 0$, these roots are
\begin{eqnarray}
\omega&=&\pm \sqrt{\gamma} c_{s,i} k - i \Sigma \frac{\gamma-1}{2\gamma}k^2 \\
\omega&=&-i\Sigma k^2/\gamma,
\label{eq:wks}
\end{eqnarray}
where the first two roots correspond to the wave propagating with the adiabatic sound speed $\sqrt{\gamma} c_{s,i}$ and the last root corresponds to the stationary mode. In the opposite limit of large $k$, the roots are
\begin{eqnarray}
\omega&=&\pm c_{s,i} k - i \frac{c_{s,i}^2 (\gamma-1)}{2\Sigma} \\
\omega&=&-i\Sigma k^2.
\label{eq:wkl}
\end{eqnarray}
In this case the propagating modes have the isothermal sound speed $c_{s,i}$.
For the wave-like structure formed at the late stage of the outburst ($t\gg t_E$), the forward propagating mode is the most relevant one for the problem at hand (see Appendix \ref{appendix:dispersion}). The propagating modes have an exponential decay time $\tau=-1/{\rm Im}[\omega]$ and the corresponding decay length $L_d\approx\tau\, d\omega/dk$. If we define the effective mean free path of electrons 
\begin{eqnarray}
l_\xi=\xi \lambda_e,
\end{eqnarray}
then the decay length becomes 
\begin{eqnarray}
L_d&=& \sqrt{\gamma} c_{s,i}  \frac{2\gamma}{\Sigma (\gamma-1)}k^{-2}\approx 0.42\; l_\xi \left ( kl_\xi\right )^{-2},~ k\rightarrow 0 \nonumber \\
L_d&=& \frac{2\Sigma}{c_{s,i} (\gamma-1)}\approx 46\; l_\xi,~ k\rightarrow \infty .
\label{eq:ld}
\end{eqnarray}
The transition between these limits corresponds to $l_\xi k\sim 0.1$. An approximation that works in both limits reads as
\begin{eqnarray}
L_d\approx 46\; l_\xi \left [1 +  8.54\times 10^{-4} (l_\xi k)^{-3}\right ]^{2/3},
\label{eq:lda}
\end{eqnarray}
which is shown in Fig.~\ref{fig:DL_dimensionless}. 
Note that the length, corresponding to energy losses of the sound wave is $\displaystyle \frac{1}{2} L_d$, since the energy of the wave is proportional to the square of the wave amplitude. With this factor the expression for $L_d$ agrees with the one in \cite{Fabian05} in the limit of small $k$, when only the effects of conduction are considered. In this limit, the phase and group velocities of the wave are close to the adiabatic sound speed $\displaystyle \sqrt{\gamma} c_{s,i}$. 

\begin{figure}
\begin{minipage}{\columnwidth}
\includegraphics[trim= 0mm 4cm 0cm 2cm, width=1\textwidth,clip=t,angle=0.,scale=0.98]{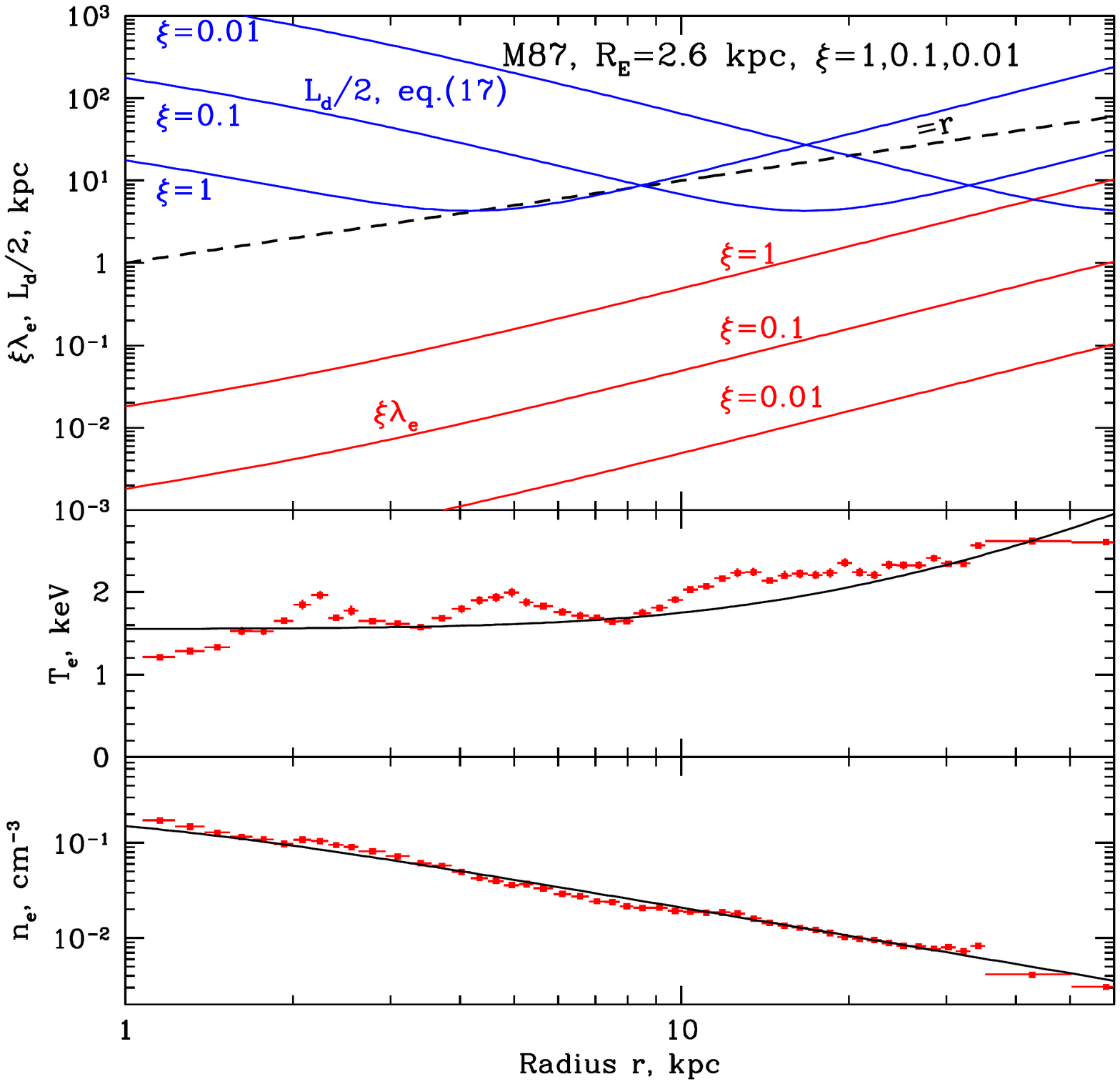}
\end{minipage}
\begin{minipage}{0.49\textwidth}
\includegraphics[trim= 0cm 4cm 0mm 2cm,width=1\textwidth,clip=t,angle=0.,scale=0.98]{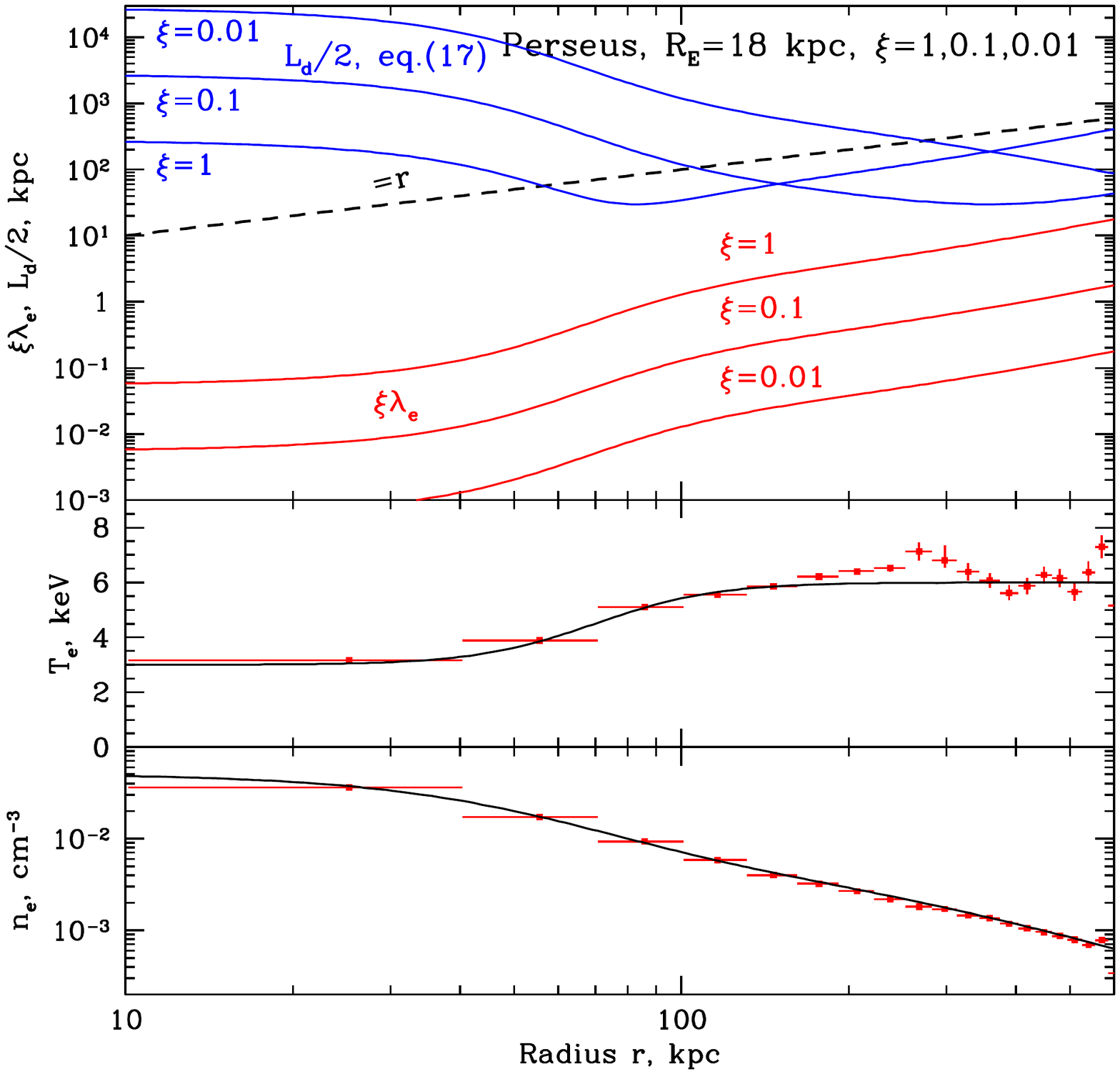}
\end{minipage}
\caption{Energy decay length ($\equiv L_d/2$) of sound waves in M87 and Perseus cluster. The top panel shows the decay length, calculated for the wavenumber $k=2\pi/3R_E$, where fiducial values of $R_E=2.6$ and 18~kpc are used for M87 and the Perseus cluster respectively. The conductivity suppression factor $\xi$ is set to 1, 0.1, 0.01 (see labels). The decay length (blue lines) is evaluated for the same value of $k$ at each radius using the local value of the mean free path (red lines). Bottom panels show the observed deprojected density and temperature profiles \citep{2007ApJ...665.1057F,2013MNRAS.435.3111Z} and the analytic approximations used to calculate $\lambda_e$. 
\label{fig:DL}
}
\end{figure}

As is clear from eq.~(\ref{eq:lda}), a formal solution of the dispersion equation (\ref{eq:disp}) predicts that the decay length is never smaller than $\approx 46\; l_\xi$ and can be much larger, if the wavenumber is smaller than $\sim 0.1 l_\xi^{-1}$.  In the limit of $k\gtrsim 0.1l_\xi^{-1}$ the wave propagates with the isothermal sound speed and has a decay length of $46\; l_\xi$. When the wavelength of the perturbation becomes comparable to the mean free path (of electrons), i.e., $k\gtrsim l_\xi^{-1}$, the diffusion limit breaks and the dissipation of sound wave depends on plasma kinetic \citep[e.g.,][]{1975PhFl...18.1287O}. The kinetic effects and ion-electron coupling have recently been discussed in \cite{Zweibel18} with a focus on A2199. In the discussion below, we, for simplicity, keep the definition of the decay length as given by eq.~(\ref{eq:ld}). As we show below (see Fig.~\ref{fig:DL}), for the relevant range of wavenumbers the waves propagating from the center are expected to decay before the isothermal regime become important. 
     
\begin{figure}
 \begin{center}
 \includegraphics[width=\columnwidth]{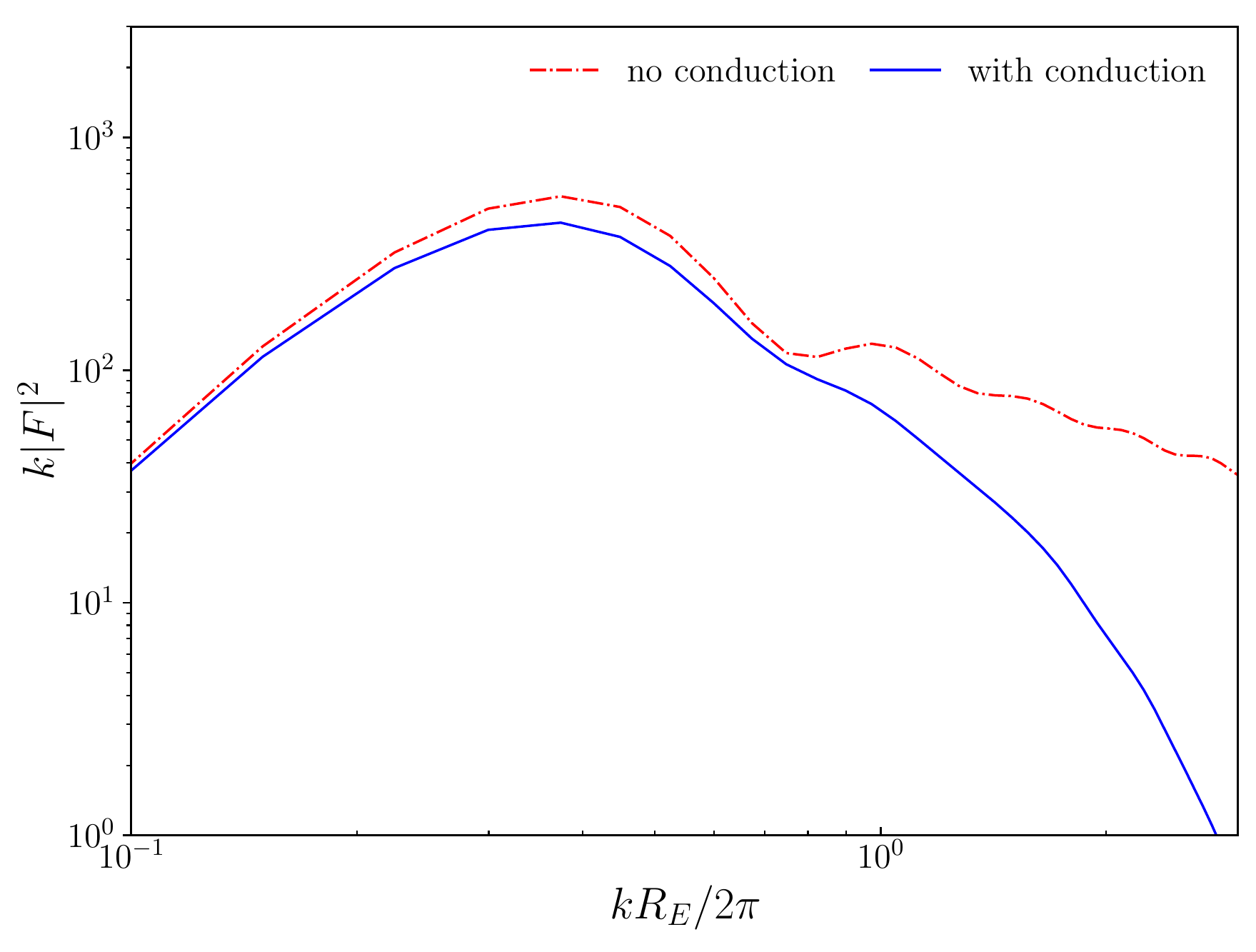} 
  \caption{Power spectrum as a function of wave number based on pressure profile at $t=10t_E$. The power spectrum $|\hat{F}|^2$ is multiplied by $k$ in order to emphasize the range of wave numbers, which contains most of the energy. Blue solid line is for the case with conduction and red dotted-dashed line denotes the case without conduction. } 
 \label{fig:power_spectrum}
 \end{center}
 \end{figure}

We can now use the arguments presented in \S\ref{sec:basic} to identify the dominant wave numbers associated with the instantaneous outburst. We did Fourier analysis of the wave like structure generated by an instantaneous outburst. The power spectrum based on pressure profile at $t=10t_{E}$ is presented in Fig \ref{fig:power_spectrum}. Similar results are also found with the density and velocity profile. This plot shows that the dominant wavelength of the outburst-driven wave is $\lambda \sim 3R_E$. When conduction is taken into account, the high frequency part is erased by the dissipation induced by thermal conduction, as they have smaller decay length. The dominant wavelength however doesn't change much and is still about $\sim 3R_E$ at least for the parameters relevant for cluster cores, even if $\xi=1$. Thus, the dominant wave number, which controls the length scale over which the weak shock decays, is $k\sim 2\pi/(\Theta_R R_E)\propto E^{-1/3}$, where $\Theta_R\sim 3$. In other words, in the frame of an instantaneous outburst model we do not have a freedom in choosing the dominant $k$, but it is defined by the total energy released during the outburst. The decay length as a function of $E$ is shown in Fig.~\ref{fig:DL} for the conditions relevant for the Perseus cluster and M87. 

\section{Discussion}
\label{sec:discussion}
Let us now discuss several possible scenarios, in the frame of the spherically symmetric outburst model characterized by the total energy release $E$ and the duration of the outburst $t_b$.

The scenario ${\mathsf L}$ is a long outburst ($t_b\gg t_E$) without conduction. In this case no strong shock is expected and the ICM is gently displaced by the growing bubble to large radii with no net heating. The energy of the outburst is split between the internal energy of the bubble $\displaystyle 1/(\gamma_b-1)P_aV$ and the work done to displace the gas $P_aV$, where $V=4\pi R_{b}^3/3$ is the volume of the bubble and $\gamma_b$ is the adiabatic index of the medium inside the bubble. The natural outcome of this scenario is the large bubble radius  $R_b\sim R_E$. The limiting case of this scenario is a quasi-continuous outburst, when the energy is supplied to the bubble at a constant rate $L_{AGN}$. In this case, the interplay between expansion of the bubble and the buoyancy forces sets the characteristic size of the forming bubble and hence the energy $E$ released during individual episodes \citep{2000A&A...356..788C}. In the simplest form, the expansion velocity of the bubble can be derived from the relation $\displaystyle 4\pi \gamma_b R_b^3P_a/3(\gamma_b-1)\sim L_{AGN}t$ and, therefore,  $\displaystyle {\rm v}_{exp}\sim R_b/t\propto L_{AGN}/(R_b^2P_a)$. At the same time, the buoyancy-induced velocity (or the velocity due to Rayleigh-Taylor instability) is $\displaystyle {\rm v}_b\sim \sqrt{gR_b}\sim {\rm v}_{k}$, where $g$ is the gravitational acceleration and ${\rm v}_k$ is the Keplerian velocity, assuming that the gravitational potential is approximately isothermal. If we assume ${\rm v}_{exp}\approx {\rm v}_b$ and $L_{AGN}=L_X$, i.e. the energy released by central AGN matches the gas cooling losses $L_X$, then the characteristic bubble radius [\citep[see eq.(5) in][]{2000A&A...356..788C}]
 \begin{eqnarray}
 &R_b \sim \left ( \frac{4}{3}\pi\frac{\gamma_b}{\gamma_b-1}{\rm v}_k \frac{P_a}{L_X}\right )^{-\frac{1}{2}}\approx  \label{eq:rb} \\ &20~{\rm kpc} \left ( \frac{{\rm v}_k}{700\;{\rm km\,s^{-1}}}\right )^{-\frac{1}{2}} \left ( \frac{P_a}{2\times 10^{-10}\;{\rm erg\,cm^{-3}}}\right )^{-\frac{1}{2}} \left ( \frac{L_X}{10^{45}\;{\rm erg\,s^{-1}}}\right )^{\frac{1}{2}}, \nonumber
 \end{eqnarray}
 which immediately sets the characteristic energy per episode
 \begin{eqnarray}
 &E_b=\frac{4}{3}\pi \frac{\gamma_b}{\gamma_b-1} R_b^3P_a \approx \label{eq:eb} \\
 &8\times 10^{59}~{\rm erg} \left ( \frac{{\rm v}_k}{700\;{\rm km\,s^{-1}}}\right )^{-\frac{3}{2}} \left ( \frac{P_a}{2\times 10^{-10}\;{\rm erg\,cm^{-3}}}\right )^{-\frac{1}{2}} \left ( \frac{L_X}{10^{45}\;{\rm erg\,s^{-1}}}\right )^{\frac{3}{2}}, \nonumber
 \end{eqnarray}
 where the adiabatic index of the ejecta $\gamma_b$ is set to $4/3$.
Thus, the energy associated with each bubble is $\sim E_b$, which is transferred to the ICM during the buoyant rise of the bubble across several pressure scale-height of the cluster atmosphere \citep[][]{2001ApJ...554..261C,2002MNRAS.332..729C,2001ASPC..250..443B}.  Neither mixing nor thermal coupling between the ICM and ejecta are involved, the ICM only ``mechanically'' interacts with the ejecta boundaries. 
     
The scenario ${\mathsf Lc}$ is the same as ${\mathsf L}$, except that the thermal conduction is included. However, if we assume that there is no heat exchange between the bubble and the ICM, i.e., the relativistic plasma inside the bubbles is thermally isolated from the ICM, then conduction is not able to tap energy from the bubble. Furthermore, the gentle expansion of the bubble doesn't induce large temperature variations in the ICM. Therefore, conduction does not make any principal impact and ${\mathsf L}$ and ${\mathsf Lc}$ scenarios are equivalent. 

Another extreme scenario ${\mathsf S}$ is a short outburst   ($t_b\ll t_E$) without conduction. In this case the shock is strong and much of the energy goes into the shock-heated gas shell with size $\sim R_E$, while the radius of the bubble becomes negligible. In addition, a sound wave carrying $\sim 12.5$\% of the outburst energy is formed at the late phases of the outburst evolution (\citeauthor{TC17}).  In the ${\mathsf S}$ scenario, there are large temperature gradients in the shock-heated shell and the outgoing sound wave also carries an appreciable amount of energy. Switching on conduction, i.e., scenario ${\mathsf Sc}$, now makes a profound impact, as the conduction can distribute the energy in the shock-heated shell and in the sound wave over larger mass of the gas, leading to net heating of the ICM. 

We note in passing that even in the absence of conduction, the shock-heated gas can be buoyant by itself and resulting convection can mix this gas with the ICM leading to the net heating. Such scenario is close to what is often implemented in numerical simulations \citep[e.g.][]{2015ApJ...815...41R}. Similarly, mixing may play a leading role if pre-heated gas is injected into the ICM.                  

Of course, the continuous energy injection scheme leading to eq.~(\ref{eq:rb}) and (\ref{eq:eb}) does not apply here, since each individual outburst is very short. However, some constraints on the energy per episode are possible. As in ${\mathsf L}$, the time-averaged energy release rate (over many outbursts) should still match the cooling luminosity $L_X$ (by virtue of the assumption of a self-regulated feedback). Also  the ${\mathsf S}$ scenario assumes that each outburst occurs in the unperturbed medium, i.e., the effect of the previous outburst is already gone. Indeed, if the outburst is happening in the preheated medium, then the shock-heating could be less efficient than in the unperturbed medium. Since the expansion of the ejecta quickly becomes subsonic with respect to the preheated plasma, much of the energy will be stored as the enthalpy of the newly forming bubble. This is plausibly the case, when the outburst is depositing energy in a bubble of relativistic plasma \citep[see, e.g.][for a discussion of Type Ia supernova outburst inside the bubble]{2011MNRAS.414..879C}. Full recovery time of the perturbed medium to the initial state can be estimated as the buoyancy time for the entire shock-heated envelope with the size $\sim R_E$. In this case, one recovers eq.~(\ref{eq:eb}) as an order of magnitude estimate, which describes the minimum energy $E_b$ needed for a single outburst. If $E_b$ is smaller, then the buoyancy is not able to remove the shock-heated gas during the typical interval $E_b/L_X$ between two successive outbursts.  The above estimates are of course qualitative and involve a number of implicit assumptions. In reality, more accurate analysis of the successive short outburst requires more specific prescription of the energy injection during individual outbursts.    
 
Of course, an actual energy release by a given AGN may not correspond to any of the limiting scenarios described above. For instance, in M87 and the Perseus cluster the values of $t_b/t_E$ are estimated to be $\sim0.2$ and $\sim0.8$, respectively (see \citeauthor{TC17} and \citealt{2007ApJ...665.1057F}). For $t_b/t_E \lesssim1$,
an outgoing sound wave, carrying  about 10-12\% of the outburst energy, is formed (see Fig. \ref{fig:energy_partition}), while the fraction of energy that goes into the shock-heated shell becomes subdominant once $t_b/t_E\gtrsim 0.2$ (see the same Figure). Therefore, for both objects we can expect that the dominant fraction of energy goes into the enthalpy of the bubble and the outgoing sound wave. In addition, M87 has smaller $t_b/t_E$ ratio, which implies the role of the shock-heated envelope is larger than for the Perseus cluster. We also note that for the Perseus cluster the best fitting models that describe the size of the bubbles, the radius and Mach number of the shock suggests that the outburst continues now \citep[see, e.g.,][TC17]{2016MNRAS.458.2902Z} and, therefore, $t_b/t_E$ ratio keeps increasing. Therefore, the ongoing outburst in the Perseus cluster could eventually evolve even close to the ${\mathsf L}$ scenario of a quasi-continuous outburst.

Finally, we note that the major caveat of the above analysis is the simplistic treatment of the conduction as the attenuated Spitzer conductivity.  
Another important simplification is associated with the assumed spherical symmetry of the outburst and the lack of momentum of the injected material. To this end, we note that the morphology of the FRI sources in the cores of many cool-core clusters suggests that the momentum of the flow is not extremely high at several kpc scales.

\section{Conclusions}
We have considered the impact of thermal conduction on the structures formed during spherically symmetric outburst of SMBH in cool-core clusters. The conduction has the strongest impact on the shock-heated shell which can form in a short outburst. For such an outburst, most of the energy is confined to this shell.  Even if the conduction is significanlty suppressed (by a factor of 10-100), it can attenuate/erase such a shell, since it is very hot, low density and compact. Therefore, the lack of the shock-heated shell might not be a good indicator for the short duration of the outburst, while the size of the central cavity is a more robust proxy for the duration of the outburst. The short and moderately long outbursts can generate a sound wave, carrying up to 12.5\% of the outburst energy. The characteristic wavelength of the outburst is set by the total energy released during the outburst. While the fraction of energy in such  wave is small, thermal conduction will help to dissipate it on scales smaller than the cooling radius, unless the conductivity is suppressed by a factor of 100 or larger. For long (quasi-continuous) outbursts all energy goes to the enthalpy of the ejecta/bubble and conduction does not play any significant role.

\section{Acknowledgements}
We are grateful to the referee -- Chris Reynolds for helpful comments.
EC is grateful to Mikhail Medvedev and Alex Schekochihin for useful discussions. 
EC acknowledges partial support by grant No. 14-22-00271 from the Russian Scientific Foundation.


\clearpage

\appendix
\section{Numerical method}\label{app:numerical}
We solve the hydrodynamic equations in the Lagrangian form
\begin{eqnarray}
\frac{\partial r}{\partial t}&=&v,\\
\frac{\partial v}{\partial t}&=& -4\pi r^2 \frac{\partial P}{\partial M},\\
\rho\frac{\partial e}{\partial t} &=&\frac{P}{\rho}\frac{\partial \rho}{\partial t} +\rho\dot{e}+\nabla \cdot \kappa\nabla T,\\
\frac{\partial}{\partial M}\left( \frac{4\pi r^3}{3}\right)&=&\frac{1}{\rho}
\end{eqnarray}
where $r$, $v$, $\rho$, $T$, $P$, $M$ and $e$ are the radius, velocity, density, temperature, pressure, Lagrangian mass and internal energy per unit mass respectively. $\kappa$ is thermal conductivity with saturation taken into account. We apply the same one dimensional hydrodynamical code as in \cite{TC17} and add a thermal conduction term in the energy equation to simulate the problem. Mass is explicitly conserved in the numerical scheme and energy conservation is maintained within $1\%$ for all the simulations.The input parameters for our default run are given in Table \ref{simulation}. 

The following scenarion is simulated. The energy is released in a few central Lagrangian cells causing the expansion of the cells boundary. The boundary drives a shock into the ICM, heating and displacing the gas (ICM). It is assumed that the central cells are thermally decoupled form the ICM, that is, the heat flux through the cell boundary is zero. Thermal conduction operates only in the gas outside the boundary. 

\begin{table}
\centering
\caption{Basic parameters for fiducial run}
\begin{tabular}{lc}
\hline\hline
injected mass $M_{ej}$& $1000 \Msun$ \\
outburst energy $E_{inj}$&$10^{59} \,\rm erg$\\
ambient electron density $n_e$& $0.01 \,\rm cm^{-3}$\\
ambient temperature $T_a$& $10^7\,\rm    K$\\
\hline\hline
\end{tabular} 
\label{simulation}
\end{table}

\section{Self similar thermal wave solution}{\label{appendix:TCD}}
For a spherical outburst in uniform medium, the thermal conduction dominated (TCD) self-similar solution \cite[e.g.][]{Barenblatt96} is available, when the expansion of shock wave driven by pressure disturbance is dynamically unimportant. The solution is characterized by the expansion of a heat front with almost isothermal interior. 

If thermal conductivity $\kappa =\kappa_{0} \rho^a T^b$, where $\rho$ is density, $T$ is temperature and $\kappa_0$ is a constant, then the spatial distribution of temperature is found to be \cite[e.g.][]{Barenblatt96}
\begin{equation}
T(R<R_h, t)=T_c\left (1-\frac{R^2}{R_h^2}\right)^{1/b},
\end{equation}
where 
\begin{equation}
T_c=\tau\varepsilon_h^{2/b}\left[ \frac{E^2(b+1)^3\rho C_v}{(\kappa_0 \rho^a)^3t^3}\right]^{1/(3b+2)}
\end{equation}
is the temperature at the center and
\begin{equation}
R_h= \varepsilon_h \left[\frac{\kappa_0 \rho^a t E^b }{(b+1)(\rho C_v)^{b+1}} \right]^{1/(3b+2)}
\end{equation}
is the heat front radius. 
\begin{equation}
\varepsilon_h= \left[2\pi \tau B\left(\frac{3}{2},\frac{b+1}{b}\right) \right]^{-b/(3b +2)}
\end{equation}
and $\tau=[b/2(b+1)(3b+2)]^{1/b}$ are two dimensionless constants, where $B(3/2,(b+1)/b)$ is the beta function. Since pressure disturbance is ignored in the evolution, the density distribution shall satisfies 
\begin{equation}
\rho(R<R_h,t)=\rho_a,
\end{equation}
where $\rho_a$ is the ambient density profile. Based on the equation of state, the pressure distribution then follows 
\begin{equation}
P(R<R_h,t)=nK_BT(R),
\end{equation}
where $n$ is the total number density.

\begin{figure}
 \begin{center}
 \includegraphics[width=\columnwidth]{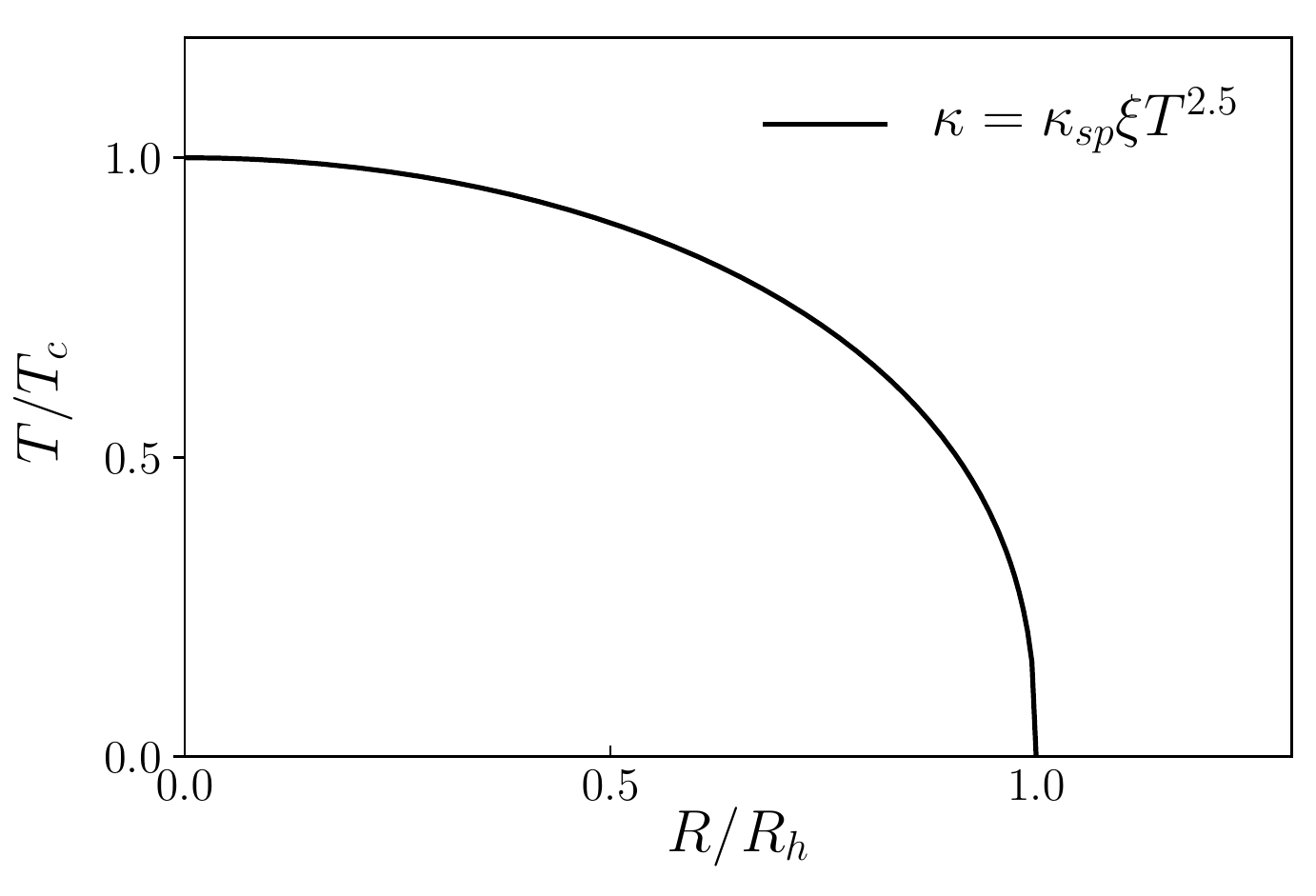} 
  \caption{Normalized TCD solution for electronic heat conduction.} 
    \label{app:fig_TCD}
 \end{center}
 \end{figure}

In AGN driven outburst, we are particularly interested in electronic heat conduction in thermal plasma with $\kappa =\kappa_{sp}\xi T^{2.5}$. $\kappa_{sp}=5\times 10^{-7}\rm erg s^{-1}cm^{-1}K^{-3.5}$ is the coefficient for the unmagnetized Spitzer conductivity, where we assume the Coulomb logrithm is 37, and $\xi$ represents the fraction of the Spitzer value. In uniform medium with $\gamma=5/3$, it is found that $\tau=0.27$ and $\varepsilon_h=1.087$. After some calculation, we obtain 
\begin{eqnarray}
T_c&=&\tau\varepsilon_h^{2/2.5}\left[ \frac{3.5^3E^2\eta_n n_e K_B}{(\kappa_{sp}\xi
)^3(\gamma-1)t^3}\right]^{1/9.5} \nonumber \\
&=&8.3\times 10^7{\rm K} ~\xi^{-6/19}\eta_n^{2/19}\left(\frac{1\rm Myr}{t}\right)^{6/19}\left(\frac{E}{10^{59}\rm erg}\right)^{4/19}\nonumber \\
&\times & ~\left(\frac{n_e}{0.01\rm cm^{-3}}\right)^{2/19}
\label{appendix:Tc}
\end{eqnarray}
and
\begin{eqnarray}
R_h&=& \varepsilon_h \left[\frac{\kappa_{sp}\xi t E^{2.5} (\gamma-1)^{3.5} }{3.5(\eta_n n_e K_B)^{3.5}} \right]^{1/9.5} \nonumber\\
&=& 19.4~{\rm kpc} ~\xi^{2/19}\eta_n^{-7/19}\left(\frac{t}{1\rm Myr}\right)^{2/19}\left(\frac{E}{10^{59}\rm erg}\right)^{5/19}\nonumber \\
&\times & ~\left(\frac{0.01\rm cm^{-3}}{n_e}\right)^{7/19}
\end{eqnarray}
where $\eta_n$ is the total number density to electron number density ratio. The temperature profile of TCD solution with electronic heat conduction is presented in Fig. \ref{app:fig_TCD} for illustration, which is characterized by an almost isothermal interior and a sharp temperature jump behind the heat front.

\section{Dispersion relation}\label{appendix:dispersion}
In this appendix, we show that how to decouple an arbitrary perturbation into three wave components, which corresponds to the three roots of $\omega$ in the dispersion relation. To make the discussion clear, we specify $\omega_1$, $\omega_2$ and $\omega_3$ for the backward propagating component, the stationary decaying component and the forward propagating component respectively.
We assume the disturbed fluid variable $X=X_0+\delta X$, where $X_0$ is the unperturbed value and $\delta X$ is the perturbed quantity. We constrain our discussion to a static uniform medium with small perturbations, i.e. $X_0$ is constant and $\delta X/X_0 \ll 1$. 
%We further define $\delta X_0(x)=\delta X(t=0,x)$. 

Let's start with one dimensional planar flow. Without loss of generality, we assume the initial perturbation satisfies  
\begin{equation}
\delta X(t=0, x)=\sum_k\delta X_{k0} e^{ikx}, 
\label{appendix: initial_condition}
\end{equation}
and the perturbation at arbitrary time $t$ is 
\begin{equation}
\delta X(t,x)=\sum_k\delta X_{k}(t) e^{ikx}, 
 \label{appendix: arbitrary_time}
\end{equation}
where $k$ is the wave number. $\delta X_{k0}$ is a constant, which indicates the amplitude of Fourier mode $k$ at $t=0$. $\delta X_{k}(t)$ is a function of time $t$, which represents the amplitude of Fourier mode $k$ at arbitrary time $t$. By definition, we have $\delta X_k(t=0)=X_{k0}$.

\begin{figure}
 \begin{center}
 \includegraphics[width=\columnwidth]{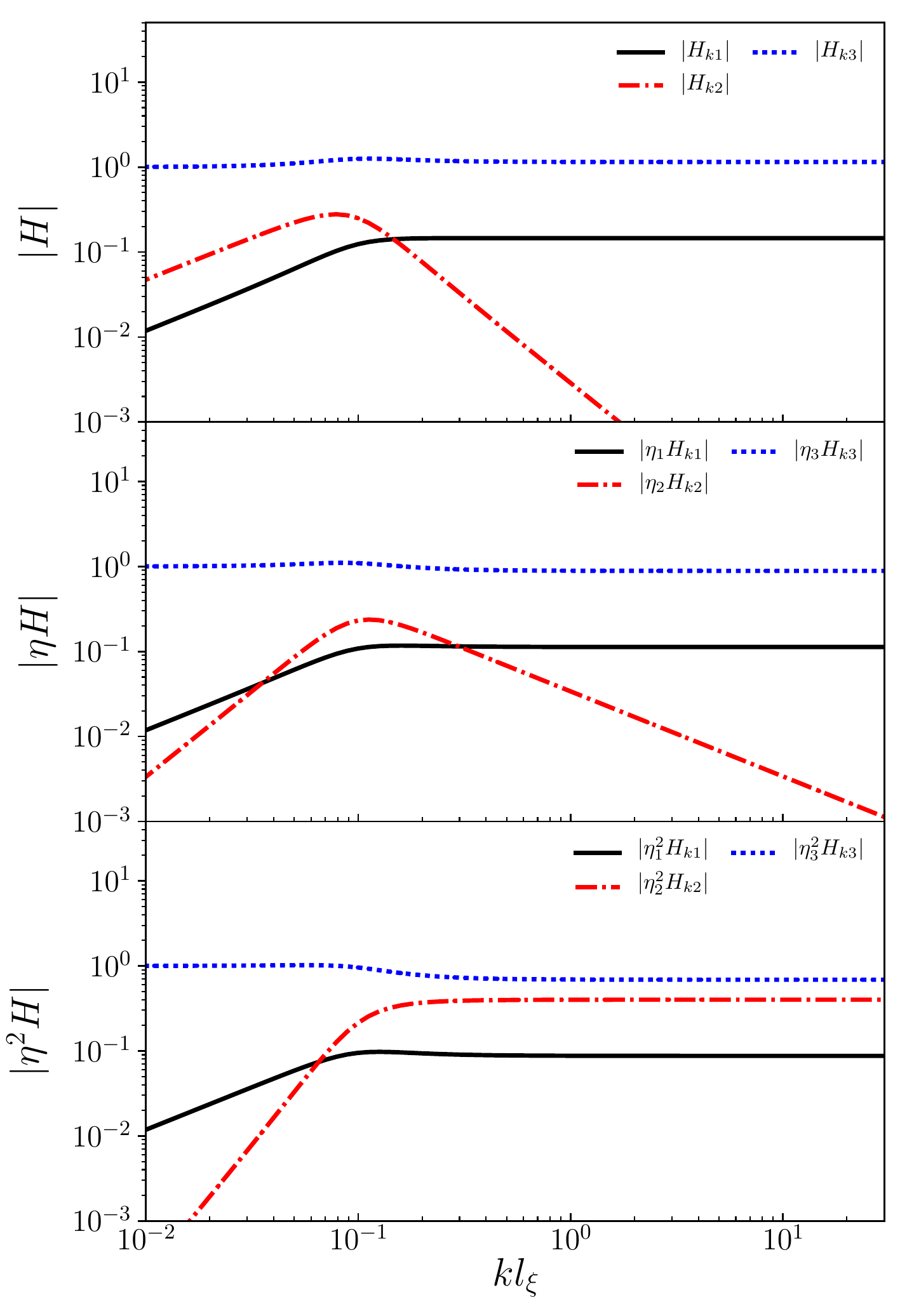} 
  \caption{Upper panel: $|H|$ as a function of $k l_\xi$, which shows the relative weight of 3 modes in perturbed density.  Middle panel: $|\eta H|$ as a function of $kl_\xi$, which indicates the relative weight of 3 modes in perturbed velocity. Lower panel: $|\eta^2 H|$ as a function of $kl_\xi$, which implies the relative weight of 3 modes in perturbed pressure.} 
    \label{app:mode_weight_fig}
 \end{center}
 \end{figure}

Each Fourier mode can be further decoupled into three wave components with angular frequency $\omega_1$, $\omega_2$ and $\omega_3$ respectively. We assume the general solution for $\delta X_k(t)$ is 
\begin{eqnarray}
\delta \rho_k(t)&=&\delta \rho_{k0}\sum_{i=1}^3 H_{ki} e^{-i\omega_{i} t}, \nonumber \\
\delta P_k(t)&=&\delta \rho_{k0}\sum_{i=1}^3 \frac{\omega_{i}^2 H_{ki}e^{-i\omega_{i} t}}{k^2}, \nonumber \\
\delta u_k(t)&=&\frac{\delta \rho_{k0}}{\rho_0}\sum_{i=1}^3 \frac{\omega_{i} H_{ki}e^{-i\omega_{i} t}}{k},
\label{app:decouple_general}
\end{eqnarray}
where $H_{k1}, H_{k2}$ and $H_{k3}$ are constants. If we further define $\eta_i=\omega_i / kc_{s,a}$, where $c_{s,a}$ is the adiabatic sound speed, then eq (\ref{app:decouple_general}) is simplified to the following form at $t=0$,
\begin{eqnarray}
1&=&\sum_{i=1}^3 H_{ki},  \nonumber \\
1&=&\frac{\delta \rho_{k0}c_{s,a}^2}{\delta P_{k0}}\sum_{i=1}^3 \eta_{i}^2 H_{ki},  \nonumber \\
1&=&\frac{\delta \rho_{k0}c_{s,a}}{\rho_0\delta u_{k0}}\sum_{i=1}^3 \eta_{i} H_{ki}.
\end{eqnarray}
By solving above 3 equations, we can derive $H_{k1}, H_{k2}$ and $H_{k3}$. The relative weight of the 3 different wave components in perturbed density, velocity and pressure are characterized by $|H_{ki}|$, $|\eta_i H_{ki}|$ and $|\eta_i^2 H_{ki}|$ respectively. For a forward propagating wave like structure with initial condition $\delta \rho_{k0} \sim \delta P_{k0}/c_{s,a}^2\sim \rho_0\delta u_{k0}/c_{s,a}$, the resulting $|H|$, $|\eta H|$ and $|\eta^2 H|$ are plotted in Fig. \ref{app:mode_weight_fig}. It is not a surprise that the perturbation is dominated by the forward propagating wave component with angular frequency $\omega_{3}$.

In one dimensional spherical outflow, we decouple the perturbed quantities into spherical waves. With this assumption, the initial condition becomes 
\begin{equation}
\delta X(t=0, r)=\sum_k\frac{\delta X_{k0} e^{ikr}}{r}.
\label{appendix: initial_condition_spherical}
\end{equation}
The perturbation at arbitrary time $t$ is 
\begin{equation}
\delta X(t,r)=\sum_k\frac{\delta X_{k}(t) e^{ikr}}{r}.
 \label{appendix: arbitrary_time_spherical}
\end{equation}
If we assume 
\begin{equation}
\delta P_k(t)\propto \delta \rho_k(t) \propto \frac{e^{-i\omega t}}{r}\, \mbox{ and }\, \delta u_k(t)\propto \frac{e^{-i\omega' t}}{r},
\end{equation} 
then the dispersion relation for one dimensional spherical flow is the same as that for one dimensional planar flow. The main difference is that $\omega'$ is no longer a constant and now depends on radius $r$. However, when $kr\gg 1$, the radial dependence becomes negligible and the discussion before about planar flow becomes a good approximation for the spherical flow. In this paper, we are interested in the decay of wave like structure in the late time evolution, where $kr \gg 1$ is satisfied for typical cluster environment and outburst energy. As a result, we can simply apply the results obtained for planar flow through our discussion. We also decouple the wave like structure generated by our numerical simulation into the three wave components with technique developed before for the planar flow. It is found that the wave like structure is indeed dominated by the forward propagating wave component, which can be used to estimate the decay length scale of wave like structure.

 % Don't change these lines
\bsp	% typesetting comment
\label{lastpage}
\end{document}